\documentclass{article}

\usepackage[english]{babel}
\usepackage{booktabs}
\usepackage[letterpaper,top=2cm,bottom=2cm,left=3cm,right=3cm,marginparwidth=1.75cm]{geometry}

\title{
The fundamental units of generalized quantum conductance and quantum diffusion
}
\author{Lino Reggiani\footnote{Dipartimento di Matematica e
Fisica, "Ennio de Giorgi", Universit\`a del Salento, via Monteroni
73100 Lecce, Italy.}, Eleonora Alfinito*, Federico Intini\footnote{Department of Engineering, "Enzo Ferrari”,
University of Modena and Reggio Emilia,
Via P. Vivarelli, 10, I-41125 Modena, Italy of Sciences and Methods for Engineering,
Via Amendola 2, Pad. Morselli - 42122 Reggio Emilia, Italy}
}
\begin{document}

\maketitle
\begin{abstract}
Although quantum transport at the nanoscale has received widespread attention since Landauer's pioneering work in 1957, we remark, that a general theory that sheds light on the difference between classical and quantum relativistic physical models is still lacking.
By considering a classical 3D gas of non-interacting quasi.particles, the article presents a unified theory that provides a generalized conductance of dimensionless quasi-particles, neutral massive, electric, thermal, and photon currents.
The investigation begins with an analogy between the original Drude model of 1900 and a modified Drude model of quasi-particles, which includes a ballistic transport regime and is independent of statistics (excluding Bose-Einstein condensation). 
Next, we construct connections between the quasi-particle unit in the modified Drude model and the carrier  unit in dimensionless, electric, massive neutral, phonon, and photon currents. 
By establishing a connection between Planck's constant $h$ and a classicaò action that takes into account the correct statistics, $h_s$, we derive the fundamental quantum unit of conductance for any of the mentioned  currents.
We further extend the diffusion coefficient of quasi-particles from the classical regime to the quantum and relativistic regimes.
We provide quantum-relativistic expressions for the generalized Einstein relations between generalized conductance and diffusion of the considered quasi-particles.
The fundamental quantum units of dimensionless, electric, neutral mass, photon/phonon energy, and the longitudinal diffusion coefficient are provided. 
Quantum relativistic expressions are given for the generalized Einstein relations between generalized conductance and diffusion of the considered quasi-particles.
It is shown that the fundamental role played by physical action is essential under both classical and quantum relativistic conditions. 
Finally, the unifying role of quasi-particle physics in connecting classical and quantum relativistic conductance is confirmed.
\end{abstract}
%
%
\section{Introduction} 
The quantization of transport phenomena at the nano-meter scale length is a well-known subject that 
dates back to Landauer pioneer formula of 1957 for the fundamental unit of  electrical quantum
conductance   \cite{landauer57,landauer88, landauer89a,landauer89b}, and subsequent contributions
by other resachers \cite{buttiker90,beenakker91,imry96,cornean05}.
Since then, quantum transport has been investigated and observed in a large number of systems such as 
the quantum Hall-effect where both electrical and thermal quantum conductance have been detected. 
In addition, quantum transport has been predicted \cite{sato05,ventra09} and also
observed in ultra-cold gases \cite{krinner15}.  
Finally, recent advances in the field of quantum heat-transport were  reviewed in \cite{pekola21}.
\par
The novelty of this  work stems  in the physical  system we consider and in the results obtained 
using an  approach that  sheds new light on the  transition  between classical and quantum or quantum 
relativistic conductance  models.
Indeed, we are referring to a sample of a two-terminal conductor
of length $L$ and cross section and contact area $A$ on which an applied driving force balanced by 
the presence of a friction, is responsible for a macroscopic steady state of the physical system. 
In particular, we choose $A  \gg L^2$ to neglect boundary effects and to be under strict 
linear-response conditions for a given temperature, $T$.
We notice that: \textit{i.} almost all the results present in the literature refer to  nano-
constraints, atomic current, or quantum point contacts; \textit{ii.} all the results make use of
 a typical  three or many-terminal structures that control the quasi-particle number, $N$, and in 
most cases exhibit a non-linear response in the applied perturbation 
\cite{kouwenhoven89,tsukada05,lambert08,milano22,marchenkov}.
\par
The central message of the paper is to have provided a theoretical development that goes beyond dimensional 
analysis and calculates new fundamental quantum units related to
neutral-mass conductance and  diffusion coefficient including quantum-relativistic quasi-particles like photons. 
Besides recovering existing outcomes concerning electrical and thermal-conductance 
\cite{greiner97,rego98,greiner00},
we want to stress that our results are not limited  to electric charge (e), but also applies to a set of 
point-like 
non-interacting  quasi-particle, (qp),  of given physical dimensions, namely: (N) their number, (e) their charge, 
(m) their neutral mass, ($\varepsilon$) their energy, like photons.
\section {Methods and results}
To begin with, we consider a classical three dimensional (3D) gas of non-interacting point-like 
quasi-particles (qp) and describe  a set of given kinetic coefficients (generalized conductance 
and diffusion coefficient), expressed in terms  of classical linear-response and diffusion laws. 
The inclusion of the ballistic transport-regime  generalizes the standard description given by the 
pioneer Drude model \cite{drude900}.
Furthermore, we investigate the analogy with quantum conditions by introducing a classical statistics 
physical-action and a classical transmission probability, as detailed in the following.
\par
By taking as significant example the electrical conductance, $G^e$, 
we introduce a modified  Drude model \cite{drude900} to include  ballistic transport-regime. 
In doing so, we define a classical transmission-probability, $\Gamma$, to obtain a modified expression of a  
quasi-particle  (qp) conductance, $G^{qp}$, as:
\begin{equation}
G^e
= 
\frac{e^2 N_e \tau} {mL^2} 
\rightarrow G^{qp}=\frac{(qp)^2 \overline{N}} {h_s}  \Gamma
\label{eq1}
\end{equation}
where in the right-hand-side (RHS)  of the above equation, the first term refers to the original Drude model 
\cite{drude900}, $e$ being the unit electric charge (in SI units), $N_e$  the particles charge-number, $\tau$
the scattering time describing  friction, $m$ the particle effective-mass. 
The second term of the RHS is the modified Drude model that applies to a non-interacting gas of point-like 
quasi-particle (qp), of given average number $\overline{N}$ (average number is associated with the open 
system nature of conductance), of physical dimension: electric charge $e$, or neutral mass $m$, 
or energy  $\varepsilon$, and introduces a statistics classical action:
\begin{equation}
h_s = m  \sqrt{\overline{v_x^{'2}}} L = k_B T \tau_T
\label{eq2}
\end{equation}
where $k_B$ is the Boltzmann constant,
$\tau_T=L/ \sqrt{\overline{v_x^{'2}}}$ is a thermal average  carrier transit-time, $
\overline{v_x^{'2}}$ is a mean squared differential (with respect to carrier number) 
quadratic velocity component along the $x$ direction \cite{gurevich79,reggiani16} given by:
\begin{equation}
\overline{v^{'2}_x} = \frac{k_BT}{m} \frac{\overline{N}}{\overline{\delta N^2}}
\end{equation}
where the bar means ensemble average over the quasi-particles  proper statistics.
Notice the explicit appearance of the Fano factor,  
\begin{equation}
\frac{\overline{\delta N^2}}{\overline{N}}
\end{equation}
to account for the effective interaction among carriers due to
the symmetry properties of their wave functions \cite{reggiani16}.
Finally, by analogy with the Landauer model \cite{landauer57}
a classical transmission probability is defined as:
\begin{equation}
\Gamma=\frac{l}{L}  
\end{equation}
with  $ 0 \leq \Gamma \leq 1$
is by definition (being $\Gamma \leq 1$) a classical transmission-probability, that describes point-like 
collisions within a scattering time $\tau$, with $l$ the associated particle mean free-path along the 
longitudinal $x$ direction given by
\begin{equation}
l=\sqrt{\overline{v^{'2}_x}} \tau
\end{equation}
The modified Drude model in Eq. (\ref{eq1}) recovers: \textit{i.} the Drude model in the diffusive 
transport regime when  $l < L$; \textit{ii.} the ballistic 1D transport regime when  $l=L$; 
\textit{iii.} the quantum limit under the quantum condition for the statistical classical action satisfying:
\begin{equation}
h_s \leq h
\label{eq2}
\end{equation}
with $h$ the fundamental Planck quantum action in SI units.
\par 
Accordingly, under the above condition $h_s$ must  be replaced by  $h$, the  Planck constant.
\par
For the ballistic 1D model, $\Gamma=1$, and the fundamental quantum unit of the ballistic
electrical-conductance reads: 
\begin{equation}
G^e_0 
= \frac{e^2 N_e \tau_T} {m L^2}  
=\frac{e^2 } {h} 
\label{eq1a}
\end{equation}
We remark, that by going from 3D to 1D, size quantization splits the original 3D energy band in a set 
of 1D mini-bands separated in energy by a quantum energy gap, 
$ \varepsilon_n$, given by
\begin{equation}
\varepsilon_n=\frac{\hbar^2 \pi^2}{2 m  W^2} n^2
\end{equation}
with $W$  the transverse-direction length and $n=1, 2, 3,...$ the sub-band index.
If not otherwise stated, here we consider only the first mini-band (also called the first transverse 
mode) by neglecting spin degeneracy so that 
$N_e=n=1$ and Eq. (\ref{eq1a}) recovers the Landauer quantum conductance unit.
\par
The \textit{electromagnetic (em) unit of electrical conductance}, $G^{em}$, is obtained by substituting 
the SI unit electric-charge with the vacuum unit-charge (Planck charge), $q_P$, as:
\begin{equation}
\frac{e^2}{q_P^2} = \alpha = \frac{e^2}{2 \varepsilon_0 h c}  
\end{equation}
with $\varepsilon_0$ the vacuum permittivity, $\alpha \simeq 1/137$ the fine structure constant and spin 
degeneracy being neglected, thus obtaining the well-known value:
\begin{equation}
G^{em}=\varepsilon_0 c
\end{equation}    
We remark that $G^{em}$ is associated with  the wave nature of the em-field, 
while  $G^e$ that is associated with the particle nature of the same field, in agreement with the dual 
property of electrical conductance \cite{reggiani16} 
\par
By analogy with the electrical conductance, by replacing the electrical potential with a chemical  
potential, the neutral-mass conductance $G^m$ for a  gas of point-like, non-interacting identical particles 
is classically given by:
\begin{equation}
G^m
=\frac{m^2 N_m} {h_s}  \Gamma
\label{eq1b}
\end{equation}
with $N_m$ the number of particle mass.
\par
For the ballistic 1D model, by analogy with the electrical conductance,  the \textit{quantum unit of 
massive conductance}, $G^m_0$, reads: 
\begin{equation}
G^m_{D0}
=\frac{m^2} {h}  
\label{eq1c}
\end{equation}
by neglecting spin degeneracy.    
\par  
For an 1D massive  electron gas the expected value of the quantum 
unit of massive conductance is estimated as 
\begin{equation}
G_0^m=1.25 \times 10^{-27} \ Kg^2/(J s)
\end{equation}
We  remark that an analogous conclusion concerning the feasibility of detecting quantized conductance in 
neutral matter was reported in \cite{lambert08,sato05} by using an analogous Landauer approach specialized 
to the flow of a gas of $^3He$ atoms dissolved in superfluid $^4He$ making use of a chemical potential as 
driving force.
\par
An experimental validation of the above equation, evidencing a quantitative step-like behavior of the massive 
conductance measured in unit of $1/h$, was carried out on a $^6Li$ neutral gas at $ T= 42 \ nK$ \cite{krinner15}.
On the same experiment we cite also  a recent theoretical speculation
 \cite{dastoor23}.
\par
For the electron thermal (\textit{th}) conductance $G^{e,th}$ we take a proper Fourier law \cite{greiner00} 
and, by analogy with the previous cases, for the 3D diffusive transport-regime we find:
\begin{equation}
G^{e,th} 
= \frac{\pi^2 N_e \tau} {3mL^2} k_B^2T 
=\frac{\pi^2 N_e k_B^2T} {3 h_s }  \Gamma
\label{eq1d}
\end{equation}
and, for the ballistic 1D model the \textit{fundamental quantum unit  of thermal conductance} for unit 
spin reads: 
\begin{equation}
G^{e,th}_0
=\frac{ \pi^2 k_B^2T} {3 h} 
\label{eq1e}
\end{equation}
in perfect agreement with similar results already published in  the literature 
\cite{greiner97,rego98,greiner00,pekola21}.
\par
For the photon radiation-conductance of a non-interacting black-body 
(\textit{bb}) photon gas, $G^{pht,bb}$, we take the analogue Fourier law  and obtain:
\begin{equation}
G^{pht,bb} 
=\frac{\pi^4  N_{ph}  k_B c} {15 \xi(3) L} 
=\frac{\pi^4  N_{ph}  k_B} {15 \xi(3) \tau_T^r} 
\label{eq1f}
\end{equation}
$N_{ph}$ being the expected non-interacting  average photon-number at thermal equilibrium in a cubic 
black-body of volume $V=L^3$, $\tau_T^r=L/c$ the relativistic transit-time, and $\xi(3)$ the Riemann 
function of the given argument.
\begin{equation}
N_{ph} = \frac{2 \xi(3)} {\pi^2 c^3 \hbar^3} (Lk_BT)^3 
\label{eq1g}
\end{equation}
Notice that the last expression of Eq. (\ref{eq1f}) is quantum-relativistic because of the presence of photons.
\par
A similar equation can be considered for the case of acoustic phonons in a solid crystal in the 
limit of temperatures below the Debye temperature value, with $s$ the sound velocity replacing $c$ in 
the definition of $\tau_T^r$ in Eq. (\ref{eq1f}) \cite{karami13}. 
\par 
By going  to diffusion, we recall the classical definition of the 
longitudinal diffusion-coefficient, $D_x$, \cite{gurevich79,reggiani16,reggiani18}:
\begin{equation}
D_x=\overline{v^{'2}_x} \tau 
= \frac{h_s}{m} \Gamma
\label{eq2}
\end{equation}
The second form of Eq. (\ref{eq2}) leads to  the quantum expression under the same conditions we used for 
the  generalized conductance.
\par
By analogy with Landauer quantum conductance model \cite{landauer57,buttiker86,imry99}, the second form of 
Eq. (\ref{eq2}) leads to define a quantum
diffusion in a gas of classical neutral particles. 
The Landauer paradigm, i.e. \textit{diffusion is transmission}, seems to be  still justified:
\begin{equation}
 D^m=\frac{h}{m} \Gamma
 \label{eq2a}
\end{equation}
For the ballistic condition, $\Gamma=1$, Eq.(\ref{eq2a})  gives the fundamental
\textit{unit of quantum diffusion} for a neutral gas of classical quasi-particles  with elementary 
 point-like mass $m$ as:
\begin{equation}
D_0^m = \frac{h}{m}
\label {eqd}
\end{equation}
A similar result was given by F\"urth in 1933 while investigating  certain relations between classical 
statistics and quantum mechanics in the framework of the Heisenberg uncertainty principle, 
as recently reported and commented in  \cite{peliti23}.
Indeed, being diffusion dimensionally the product between length and  velocity,  Eq. (\ref{eqd}) 
represents the uncertainty relation between a particle position and velocity in both the cases of quantum 
(when the mass takes its classical rest-value)  and relativistic  (when the mass takes its energy equivalent 
value, $m_{r}=hf/c^2$, with $f$ the photon frequency).
We would stress that to define a quantum expression for the diffusion coefficient, since $h$ cannot  
be factorized,  we need to introduce the mass of the quasi-particle, contrary to  the classical  or 
relativistic cases  where velocity and length or wavelength are sufficient.
Indeed, from a dimensional point of view quantum diffusion is action/mass, while for classical or 
relativistic diffusion is  velocity $\times$ space.
\par 
For an electron gas $D _0^m= 7.3 \times 10^{-4} \ m^2/ s$ that, in the absence of experiments, can be 
compared  with the experimental values in Si at $77 \ K$ of $1.6 \times 10^{-2}\ m^2/ s$  and of
 $0.61 \times 10^{-2}\ m^2/ s$, and  at $300 \ K$ of $3.5 \times 10^{-3}\ m^2/ s$  and of $1.5 \times 
10^{-3}\ m^2/ s$ for electrons and holes, respectively under classical diffusive regime of electrical transport 
\cite{jacoboni83,reggiani85}. 
\par                                
Interesting enough, the single particle mass satisfies the kinetic expression:
\begin{equation}
 D_0^m G_0^m =  m 
\end{equation}
that, as we shall see in the following,  is valid in both classical and quantum/relativistic  cases.
In addition, the classical expression of the  generalized Einstein relation \cite{einstein05}
\begin{equation}
G^e =  \frac{e^2 m \overline{N_e}}{(Lm)^2 \overline{v_x}^{'2}} D^m
= \frac{e^2 m \overline{N_e}}{h_s^2} D^m
\end{equation}
%
for the fundamental quantum unit (i.e. $\Gamma$ and $\overline{N_e}=1$, $h_s = h$) gives:
\begin{equation}
 G_0^e =  \frac{e^2 m}{h^2} D_0^m 
\end{equation}
We further notice that the classical and quantum  definition of diffusion can be extended to full
relativistic particles, that is for a  monochromatic photon gas \cite{reggiani18} the 
quantum relativistic diffusion, $D_0^{qr}$, is given by:
\begin{equation}
 D_0 ^{qr}  = \frac{h}{m_r} = c \lambda
\label{edr}
\end{equation}
Notice that the last equality in Eq. (\ref{edr})  is the fundamental unit of the quantum relativistic 
diffusion in terms of the photon wavelength $\lambda$.
\par
By considering the black-body conductance radiation, $G^{pht,bb}$, the quantum relativistic case associated  
with photon conductance in the presence of a temperature gradient for a macroscopic volume 
$L^3$  is \cite{karami13}:
\begin{equation}
G^{ph} = \frac{\pi^5 }{15 \zeta(3)}\frac{k_B \overline{N_p} 
D_0^{rel}}{L^2}
\label {2eq}
\end{equation}
with the average photon number being
\begin{equation}
\overline{N_p}   = \frac{2^4 \pi \zeta(3) L^3}{(hc)^3} (k_B T)^3
\end{equation}
$\zeta(3)$ being the associated Riemann function.
We notice that being photon already a quantum relativistic  quasi-particle, in
the absence of photon interaction and  due to  their velocity $c$  the radiation conductance is macroscopic 
and ballistic also in 3D. 
\par
Equation (\ref{2eq})  represents the quantum relativistic  generalized Einstein relation between 
radiation conductance and diffusion. 
By analogy with other fundamental units of conductance, 
the fundamental unit of quantum radiation-conductance can be taken for the case of $\overline{N_p} =1$ as:
\begin{equation}
G_0^{ph} =\frac{2 \pi^5 }{15 \zeta(3)} k_B\frac{c}{L}
\end{equation}
\section{Discussion}
The results obtained in the paper are summarized in the following four tables.
\par  
Table I reports the expressions of the diffusion coefficient associated with different physical 
approaches (classical, relativistic, quantum  and quantum  
relativistic) and different transport regimes (diffusive and ballistic).
\par
Table II reports the  quasi-particles (number, electrical, neutral massive, energy) involved in the 
fundamental unit of quantum diffusion and conductance  available from literature and obtained in the 
present paper by applying the uncertainty  principle to the classical expressions obtained by an 
appropriate modification of the original Drude classical model for the electrical property of metals.
\par 
Table  III reports the first quantum-generalized Einstein relation between conductance and diffusion 
(i.e. G/D) for the different quasi-particles considered here.
\par 
Table  IV reports the second quantum generalized Einstein relation between conductance and 
diffusion (i.e. $G \times D$) of the different quasi-particles considered here.
Remarkably, results show that quantum effects are washed out in the second generalized Einstein 
relations.  
In other words, the product between conductance and diffusion is found to be independent  of $h$.
\par
In a first comment we notice that each 1D fundamental quantum unit
contains three basic physical quantities: the concept of a transverse mode 
replacing the number of quasi particles (when degeneracies due to  spin or polarization 
properties are neglected), the Planck universal constant, as signature of quantum mechanics or more generally
of the importance of the physical action, and the dimension of the physical quasi-particle of interest.  
In the present case these can be the number, the charge, the mass and the energy of the quasi-
particle, and the vacuum. We remark, that going from classical to quantum conditions we consider a 
single transverse energy mode by neglecting spin or polarization degeneracies, that can be responsible 
for a factor of two for electrons or photons, respectively.     
In the case of vacuum, electro-magnetic conductance is a property of vacuum electrical permittivity 
and magnetic permeability, as dictated by Maxwell equations. 
\par
In a second comment we notice that the ballistic  condition implies  macroscopic quantities that are 
expressed in a global  form  that cannot be factorized in a geometrical and a local physical contribution.
As typical example, the local conductivity responsible of friction becomes dependent of the sample 
length because the scattering time should be replaced by the transit time, $\tau_T = 
L/\sqrt{\overline{v_x^{'2}}}$.  
The introduction of a classical transmission probability, $\Gamma$,  
enables us to include these two fundamental representations of a transport phenomenon (identified 
as diffusion and conductance) under the condition $0 \leq  \Gamma \leq 1$, that justifies the probabilistic 
interpretation of $\Gamma$.
Following Landauer (1957), in a consistent quantum approach, $\Gamma$ should be evaluated using a 
quantum theory of scattering.  
However, in the present case the ballistic transport regime is of 
importance, that implies $\Gamma=1$ in both classical  and quantum approaches, otherwise, due to 
the dependence from the interaction, the universality of the results will be lost. 
However, we want to stress that by introducing the classical transmission probability, local scatterings  
are no longer necessary to define a diffusion and/or a conductance, since the role of
the contacts by defining the length of the sample suffices. 
Indeed, the inclusion of local scatterings, besides making the results no longer universal,
has the net effect to  decrease the maximum ballistic value of  conductance down to zero for a 
vanishing value of $\Gamma$.
In other words, we want to emphasize the crucial importance of introducing a transmission probability 
already in the classical case (see Eqs. (\ref{eq1}, \ref{eq1b})) that enables us to 
account for the presence of the terminals independently of  the point-like scattering time 
$\tau$ \cite{reggiani23}.

\begin{table}[pt]
\centering
\caption{
Diffusion expressions under different physics and transport regimes.
} 
\vskip 2pt
{\begin{tabular}{@{}ccc@{}} \toprule \\
	   Diffusion $D_x$ & Physics & Transport regime 
		\\ \\
		\hline
		\\
	$\sqrt{\overline{v^{'2}_x}} l$ & Classical    & Diffusive \\	
	$\sqrt{\overline{v^{'2}_x}} L$ & Classical    & Ballistic \\	 
	$c L$ & Relativistic    &  Ballistic \\ 
	$\frac{h}{m} \Gamma $ & Quantum    &  Diffusive \\	
	$\frac{h}{m}$ & Quantum    &  Ballistic \\	
  $c \lambda$ & Quantum  relativistic   & --- \\	
	\\
	\hline
		\end{tabular}}
\label{tab:fununi}
\end{table}
%

\begin{table}[pt]
\centering
\caption{
Diffusion and  universal units of quantum conductance for a 1D ballistic conductor of length $L$ 
for a  quantum gas of non-interacting  quasi-particle (qp) of given spin/polarization, number (N), 
massive (m), electrical (e), single photon-energy ($\varepsilon$), thermal carrier-energy (th), 
black-body photon-energy (pht), phonon-energy  (phn), electro-magnetic (em). Here $h$ is the 
Planck constant, $c$ the light velocity in vacuum, $m_rc^2=\varepsilon$ the relativistic equivalent  
mass-energy, $k_B$ the Boltzmann constant, $T$ the absolute temperature, $s$ an average acoustic
 phonon velocity, $\varepsilon_0$ the vacuum permittivity.
} 
\vskip 2pt
{\begin{tabular}{@{}ccc@{}} \toprule \\
	  qp & Diffusion $D_0^{qp}$ & Conductance $G_0^{qp}$  
		\\ \\
		\hline
		\\
	$N$ & $\frac{h}{m} $   & $\frac{1}{h} $ \\		
	$e$ & $\frac{h}{m} $   & $\frac{e^2}{h} $ \\	
 	$m$ & $\frac{h}{m}$    & $\frac{m^2}{h}$   \\
  $\varepsilon$ & $\frac{h}{m_r} $   & $\frac{\varepsilon^2}{h} $ \\	
	$th$ & $\frac{h}{m}$   &$\frac{\pi^2}{3} k_B\frac{k_B T}{h}$ \\ 
	$pht$ & $cL$  &$\frac{\pi^5}{15 \zeta(3)}k_B\frac{c}{L}$  \\ 
  $phn$ & $sL$  &$\frac{\pi^5}{15 \zeta(3)}k_B\frac{s}{L}$  \\ 
	$em-vacuum$	& $cL$  &$\varepsilon_0 c$  \\
	\\
	\hline
		\end{tabular}}
\label{tab:ger}
\end{table}
%

\begin{table}[pt]
\centering
\caption{
First generalized Einstein relations of conductance diffusion for quantum fundamental units 
of 1D ballistic conductors  of length $L$ for a single quantum quasi-particle (qp) of given s
pin/polarization (number (N), massive (m), electrical (e), single photon-energy ($\varepsilon$), 
thermal-energy (th), photon-energy (pht),  phonon-energy (phn) and (em-vacuum)).
}
\vskip 2pt
{\begin{tabular}{@{}ccc@{}} \toprule \\
	  qp & First quantum Einstein relation  (G/D)
		\\  \\
		\hline
		\\
	$N$ & $\frac{G_0^N}{D_0^N}=\frac{m}{h^2}$  \\	
	$e$ & $\frac{G_0^e}{D_0^e}=\frac{e^2m}{h^2}$  \\
	$m$ & $\frac{G_0^m}{D_0^m}=\frac{m^3}{h^2}$    \\
	$\varepsilon$ & 
	$\frac{G_0^{\varepsilon}}{D_0^{\varepsilon}}=
	\frac{\varepsilon}{\lambda^2}$  \\		
$th$ & $\frac{G_0^{th}}{D_0^{m}}=\frac{\pi^2}{3} \frac{k_B^2 T}{h^2}$ \\
$pht$ & $\frac{G_0^{pht}}{D_0^{th}}=\frac{\pi^5}{15 \zeta(3)}\frac{k_B^2T}{h^2} $  \\
$phn$ & $\frac{G_0^{phn}}{D_0^{th}}=\frac{\pi^5}{15 \zeta(3)}\frac{k_B^2T}{h^2}$  \\
$em-vacuum$ & $\frac{G_{vacuum}}{D_{vacuum}}=\frac{\varepsilon_0}{L}$  \\
	\\
	\hline
		\end{tabular}}
\label{tab:ger}
\end{table}
%
\begin{table}[pt]
\centering
\caption{
Second generalized Einstein relations of conductance diffusion for quantum fundamental 
units of 1D ballistic conductors  of length $L$ for a single quantum quasi-particle (qp) of 
given spin/polarization (massive (m), electrical (e), thermal-energy (th), photon-energy (pht)
and vacuum, with  $\mu_0$ being the  magnetic permeability.
}
\vskip 2pt
{\begin{tabular}{@{}ccc@{}} \toprule \\

  qp & Second quantum Einstein relation  ($GD$)
		\\ \\
		\hline \\
	$N$ & ${G_0^N} \times {D_0^N}=\frac{1}{m}$  \\		
	$e$ & ${G_0^e} \times {D_0^e}=\frac{e^2}{m}$  \\		
	$m$ & ${G_0^m} \times {D_0^m}=m$    \\
  $\varepsilon$ & ${G_0^{\varepsilon}} \times {D_0^{\varepsilon}}=\varepsilon c^2$  \\	
  $th$ & ${G_0^{th}} \times {D_0^{m}}=\frac{\pi^2}{3} k_B\frac{k_BT}{m}$   \\
  $pht$ & ${G_0^{pht}} \times {D_0^{th}}=\frac{\pi^5}{15 \zeta(3)}
	k_B c^2$  \\
  $phn$ & ${G_0^{phn}} \times {D_0^{th}}=\frac{\pi^5}{15 \zeta(3)} 
	k_B s^2$  \\	
	$em-vacuum$ & $G_{vacuum} \times D_{vacuum}=\frac{L}{\mu_0}$  \\
	\\
	\hline	
		\end{tabular}}
\label{tab:ger}
\end{table}
\section{Conclusions}
The paper develops  a unitary physical approach that starts from a classical phenomenological kinetic-laws 
in a 3D sample, now given by a modified Drude law.
Then,  in the limit of a 1D geometry and a ballistic transport-regime,
the  introduction of appropriate quantization conditions 
enables us to formulate a set of fundamental quantum units of conductance and diffusion
going beyond the pioneer Landauer approach limited electrical conductance. 
Specifically, we complete the quantum definition of available  fundamental kinetic coefficients associated with 
linear response \cite{greiner00,pekola21}; i.e. number, electrical, neutral mass, 
thermal/photon/phonon-energy, vacuum conductance, and diffusion by using a unified physical approach.
Theory is validated by recovering existing results in the literature.
\par
An alternative form of the quantum generalized Einstein relation
and quantum generalized mass conductance diffusion evidences an intriguing property of the product 
conductance-diffusion that is found to be compatible with the classical results. 
Thus,  the Landauer paradigm that conductance and diffusion are transmission  is here extended to 
classical conditions.
\par
We finally would like to comment on the concept of conductance.   I
ts definition is usually related to a model of driving force and friction applied to a single quasi-particle 
in the presence of a constant velocity whose balance gives the conductance.  However, conductance of a gas of 
quasi-particles can also be defined in terms of fluctuation dissipation relations at thermal equilibrium 
(see \cite{greiner00}, thus in the absence of both: an external driven force and a friction.  
Furthermore, also in the ballistic transport regime, again conductance can be defined in the absence of 
driven force and friction, but in the presence of an action (classical or quantum) associated with the quasi-
particle.
This includes the case of a monochromatic photon gas where by definition any quasi-particle moves at constant 
velocity $c$ in vacuum.
Here, the fundamental unit of quantum-relativistic photon  energy-conductance can be expressed in 
several equivalent forms as:
\begin{equation}
G_0^{\varepsilon} = \frac{ \varepsilon^2}{h} =  \frac{ m_r^2 c^4}{h}  =  \varepsilon f =  hf^2
\end{equation}
\section{Acknowledgments}
Prof. Dr. Tilmann Kuhn from M\"unster University is warmly thanked for the very valuable comments he provided on the  subject. 
Prof. Dr. Luca Peliti,  Deputy Director Santa Marinella Research Institute (Roma) is acknowledged for valuable comments on the quantum diffusion model
and  for providing us with a copy of Ref. \cite{peliti23}.
\section{Conflicts of interests}
All  authors declare that they have no conflicts of interest.
\section{Data availability statement}
All data that support the findings of this study are included within the article.
\section{Funding}
Authors did not receive specific funding
for this research
\end{document}